\documentstyle[12pt,A4wide]{article}
\begin{document}
\rightline{DFPD 94/TH/24, May 1994}
\rightline{DF UPG 87/94, May 1994}

{\begin{center}
{\Large \bf Classical and quantum dissipation} \\
\vspace{0.3cm}
{\Large \bf in non homogeneous environments} \\
\vspace{0.8cm}
{\large Fabrizio Illuminati\footnote{Bitnet address:
illuminati@mvxpd5.pd.infn.it}} \\
\vspace{0.1cm}
{\em Dipartimento di Fisica ``Galileo Galilei", \\
     Universit\`a di Padova, and INFN di Padova, \\
     Via F. Marzolo 8, 35131 Padova, Italia} \\
\vspace{0.6cm}
{\large Marco Patriarca\footnote{Bitnet address:
patriarca@perugia.infn.it}} and
{\large Pasquale Sodano\footnote{Bitnet
address: sodano@perugia.infn.it}} \\
\vspace{0.1cm}
{\em Dipartimento di Fisica, Universit\`{a} di Perugia, \\
and INFN di Perugia, Via A. Pascoli, 06100 Perugia, Italia} \\

\vspace{1.3cm}

{\large \bf Abstract}

\end{center}}

We generalize the oscillator model of a particle interacting
with a thermal reservoir by introducing arbitrary nonlinear
couplings in the particle coordinates.
The equilibrium positions of the
heat bath oscillators
are promoted to space-time functions, which
are shown to represent a modulation
of the internal noise by the external forces.
The model thus provides a
description of classical and quantum dissipation in
non homogeneous environments.
In the classical case we derive a generalized Langevin
equation with nonlinear multiplicative noise and a
position-dependent fluctuation-dissipation theorem associated
to non homogeneous dissipative forces.
When time-modulation of the noise is present,
a new force term is predicted
besides the dissipative and random ones.
The model is quantized to obtain the
non homogenous influence functional and master equation
for the reduced density matrix of the Brownian particle.
The quantum evolution equations reproduce
the correct Langevin dynamics
in the semiclassical limit.
The consequences for the issues of decoherence and localization
are discussed.

\vspace{0.6cm}


\newpage

{\large \bf 1. Introduction and summary.}

\vspace{0.5cm}

In the Langevin approach to classical Brownian motion \cite{Van81}
the total system, globally isolated, is divided into two parts, the
central particle and the environment. The effects of the environment
on the otherwise free particle are taken into account by adding
two force terms in Newton's equation, a deterministic damping force
$-M\gamma_{\scriptscriptstyle 0}\dot{x}$ and a random force $R(t)$
(for simplicity
we consider only one-dimensional motions, since the generalization
to higher dimensions is straightforward):
\begin{equation}
M\ddot{x} = - M\gamma_{\scriptscriptstyle 0}\dot{x} + R(t).
\end{equation}

The former expression is the
Langevin equation, where $M$ stands for
the mass of the Brownian particle, $\gamma_{\scriptscriptstyle 0}$
is the damping
constant, and $R(t)$ is a
stationary Gaussian Markov process.
The expectation value $ \left\langle R(t) \right\rangle $ of
the random force is assumed to be zero,
\begin{equation}
\left\langle R(t) \right\rangle = 0,
\end{equation}

\noindent so that the macroscopic equation for the expectation
$\langle x(t)\rangle $ coincides with the classical
equation of motion of
a damped particle, $M\ddot{x}_{cl} = -M\gamma_{\scriptscriptstyle 0}
\dot{x}_{cl}$.
The additional requirement that, for $t\rightarrow \infty$,
the particle reaches thermal equilibrium with the environment,
so that $M\langle \dot{x}^{2} \rangle /2
= k_{B}T/2$, requires that:
\begin{equation}
\left\langle R(t)R(s) \right\rangle =
2\gamma_{\scriptscriptstyle 0}Mk_{B}T\delta(t-s),
\end{equation}

\noindent where $k_{B}$ is Boltzmann's constant and $T$ is the
absolute temperature. $R(t)$ is thus a white noise.
Eq.(3) expresses the well known fluctuation-dissipation theorem
for the particular case of white noise, in which
the correlation function $K(t-s) \propto
\delta(t-s)$.

However, there is no Lagrangian and therefore no action
principle allowing to derive the phenomenological
equation of classical Brownian motion, eq.(1).
This shortcoming of the
classical theory makes it very difficult to
build a quantum theory of Brownian motion and to treat the
quantum mechanics of open systems
because canonical or
path-integral quantization require the knowledge of the
microscopic dynamics of the system, either
Lagrangian or Hamiltonian.

A possible way to overcome this difficulty is to introduce a
suitable mechanical model of dissipation. A particular one,
the so-called oscillator model, has grown to
a privileged position since its inception
in the early sixties
(\cite{Fey63}, \cite{Ford65}, \cite{Ull66}).
The oscillator model, in its
classical version, provides an appealing alternative derivation
of the Langevin equation of classical Brownian motion.
Upon quantization, the model provides a theory of quantum
Brownian motion which, in the semiclassical limit,
reduces to the known phenomenological Langevin dynamics.

The first step towards a mechanical model of
homogeneous classical and
quantum dissipation is to consider the central particle and
the dissipative environment as forming a large isolated dynamical
system.

The second step is to consider the dissipative environment
as composed by an infinite collection of
independent harmonic oscillators $ \{ q_{n} \} $, with frequencies
$ \{ \omega_{n}/2\pi \} $, linearly
coupled to the ``central" particle $x$ of mass $M$.
Then, if suitable assumptions are made on the statistical
distribution of the initial conditions of the oscillators,
the central particle exhibits Brownian motion
with homogeneous noise in the limit
of continously distributed frequencies.

The Lagrangian of this many-body mechanical model is
\begin{equation}
L(x,\dot{x}, \{ q_{n},\dot{q}_{n} \} ,t)=\frac{M}{2}\dot{x}^{2}
+  \sum_{n} \left\{ \frac{m}{2}\dot{q}_{n}^{2} -
\frac{m}{2}\omega_{n}^{2}(q_{n} - x)^{2} \right\},
\end{equation}

\noindent where for simplicity all the oscillators are assumed
to be of equal mass $m$.
The first term in the r.h.s. of eq.(4) is the free
Lagrangian $L_{0}$
of the central particle, while the remaining part will be
referred to as the {\em oscillator} Lagrangian $L_{osc}$;
it describes the {\em whole} environment and its interaction
with the central particle.
The oscillator Lagrangian actually describes a field-particle
coupling, for any free field can be essentially viewed as an
infinite collection of harmonic oscillators.

The oscillators are assumed to be
in thermal equilibrium at the initial time $t_{0}$
according to a canonical distribution:
\begin{equation}
\begin{array}{lll}
\langle q_{i0} - x_{0} \rangle & = & 0 \, , \\
&& \\
\langle \dot{q}_{i0} \rangle & = & 0 \, , \\
&& \\
m\omega_{i}\omega_{j} \langle
(q_{i0}-x_{0})(q_{j0}-x_{0}) \rangle & = & k_{B}T \delta_{ij} \, , \\
&& \\
m\langle \dot{q}_{i0}
\dot{q}_{j0}\rangle & = & k_{B}T \delta_{ij} \, , \\
&& \\
\langle \dot{q}_{i0}(q_{j0} - x_{0}) \rangle & = & 0 \, ,
\end{array}
\end{equation}

\noindent where $\delta_{ij}$ is the Kronecker symbol and
$q_{i0}=q_{i}(t_{0})$, $\dot{q}_{i0}=\dot{q}_{i}(t_{0})$, and
$x_{0}=x(t_{0})$.
The classical homogeneous Brownian dynamics of the central particle
is obtained by
eliminating the oscillators coordinates from the Euler-Lagrange
equations.
The resulting equation of motion for the central particle $x$ is the
generalized Langevin equation with homogeneous
noise, considered by Mori and Kubo in the context of nonequilibrium
statistical mechanics \cite{Mori65},
\begin{equation}
M\ddot{x} = - \int_{t_0}^{t}dsK(t-s)\dot{x}(s) + R(t).
\end{equation}

The homogeneous noise $R(t)$ and the
homogeneous memory kernel
$K(\tau)$ are, respectively,
\begin{eqnarray}
R(t) & = & \sum_{n} m\omega_{n}^{2} \left\{ (q_{n0}-x_{0})
\cos [ \omega_{n}(t-t_{0}) ]
+ \frac{\dot{q}_{n0}}{\omega_{n}}\sin [ \omega_{n}
(t-t_{0}) ] \right\} , \\
&& \nonumber \\
K(\tau) & = & \sum_{n} m\omega_{n}^{2}\cos{(\omega_{n}\tau)}.
\end{eqnarray}

\noindent Note that the initial oscillator coordinates and
velocities are just
the Fourier coefficients of $R(t)$.
By using initial conditions (5), it is straightforward to show that
$R(t)$ is connected to the memory kernel $K(\tau)$ by the
fluctuation-dissipation theorem:
\begin{equation}
\left\langle R(t)R(s) \right\rangle = k_{B}TK(t-s).
\end{equation}

The form of the correlation function depends on the oscillator
distribution. For instance, white noise
corresponds to an oscillator distribution function
$G(\omega)=2M\gamma_{\scriptscriptstyle 0}/\pi m \omega^{2}$.
In this case the
correlation function is
\begin{equation}
K(\tau) = \int_{0}^{\infty}d\omega G(\omega)m\omega^{2}
\cos{(\omega \tau)} = 2\gamma_{\scriptscriptstyle 0}\delta(\tau)
\end{equation}

\noindent and eq.(6) and (9) reduce to eq.(1) and (3) respectively.

It should be remarked that the dissipation-fluctuation theorem
was originally obtained on grounds of internal consistency
of the Langevin equation \cite{Mori65}, while in the
oscillator model it is a natural consequence of the form of the
Lagrangian (4) and of the initial conditions (5).

The model Lagrangian (4)
is the only quadratic one which is both
reflection and translation invariant. These invariance
properties must be shared by any model Lagrangian describing
a globally isolated system, in order for eq.(1) to be invariant.

In fact, the homogeneous oscillator model can be effectively
visualized as a set of masses $m$ bound to the mass $M$ of the
central particle by linear forces \cite{Gra88}.
In particular, the coordinate $x$ of the
central particle represents the equilibrium position of the
oscillators. This explains the presence of $x_{0}$ in the
initial data of the oscillators, eq.(5).
Lagrangian (4) has been first proposed in
ref. \cite{Ford65}, and it
has been later considered in different contexts (\cite{Zwa73},
\cite{Hak85}, \cite{Ford87}, \cite{Sch87}).

Models with an oscillator Lagrangian which is not translation
and reflection invariant
produce non physical infinite force terms and ill-defined
quantities which depend on frequency cutoffs, both in the
classical and in the quantum regime (\cite{Ull66},\cite{Cal83},
\cite{Eck87},\cite{Hu92}).

Consider now the case in which an external potential $V(x,t)$
is present. The usual generalization of eq.(6) is
\begin{equation}
M\ddot{x} + \partial_{x}V(x,t)
= - \int_{t_{0}}^{t}dsK(t-s)\dot{x}(s) + R(t),
\end{equation}

\noindent obtained by simply adding the external force $-\partial_{x}
V(x,t)\equiv \partial V(x,t)/ \partial x$ in the generalized Langevin
equation.
It can be derived from the
oscillator model by simply
adding the potential term $V(x,t)$ in Lagrangian (4).
In many physical situations this equation or
its white noise limit provide a satisfactory
description of classical Brownian dynamics.

However, from a general point of view
some inconsistencies arise in this approach.

At a phenomenological level
one can see, by inspection of eq.(11), that the addition
of an external potential $V(x,t)$ has affected the mechanical
part of eq.(6), on the left hand side, leaving unchanged the
environment-induced force terms, on the right hand side.
Instead, a corresponding influence should be expected to
appear also on these terms, since the environment
surrounding the Brownian particle is made up of particles
as well.
Thus, eq.(11) cannot be in general a correct
description of Brownian dynamics in the presence of
external fields.

There are many physical situations in which the non
homogeneous nature of the environment has observable consequences
on the dynamics of the central degree of freedom. Some examples are
provided by systems escaping an oscillating barrier \cite{Doe92},
stochastic resonance
or stochastic parametric oscillators \cite{Stra81}.

The purpose of the present paper is to provide a reliable model
for the description of classical and quantum Brownian motion
in non homogeneous environments.
This is achieved by introducing
a suitable generalization of the classical homogeneous
oscillator model.

A true mathematical modellization
of the bath and the central particle as forming a closed system
is really sound as long as $V(x,t)=0$. In this case $L_{0}$
and $L_{osc}$, as well as the total Lagrangian, share the
same invariance properties.
If we allow for external potentials $V(x,t)$ acting on the central
particle, we are in fact considering situations in which the whole
system (bath + particle) is open and it is then
natural to consider it coupled to the rest of the universe
through external forces which act both on the particle {\em
and} on the thermal oscillators. In this way one treats
the central particle and the oscillators on equal dynamical footing.

This approach enables one to study those physical situations where
strong enough couplings with external fields do not leave
the environment unaffected.

The generalized Lagrangian is introduced in Sec.2.
Besides adding an external potential $V(x,t)$,
we let external forces $F_{n}(x,t)$ act upon each
oscillator $q_{n}$ with coupling strength $c_{n}$, in order
to describe the action of the external potential on the environment.
This is shown to be equivalent to shifting the equilibrium positions
of the oscillators from $x$, the position of
the central particle, to the positions
$Q_{n}(x,t) = c_{n}F_{n}(x,t)/m\omega_{n}^{2}$. It is also shown that
the functions $Q_{n}$ represent a space-time modulation of noise
and dissipation.

We treat the classical model first, and for it we derive the
Euler-Lagrange equation of motion for the central particle.
It turns out to be a modified generalized Langevin equation with a
colored and multiplicative noise term $\tilde{R}(x,t)$ that can be
expressed as the gradient of a stochastic potential $\tilde{V}(x,t)$.
Further, a position-dependent memory kernel $\tilde{K}$ replaces the
homogeneous kernel $K(\tau)$ in the dissipative force term.
A new deterministic force is predicted besides
the random and dissipative ones when time-modulation of the
noise is present.
A generalized position-dependent
fluctuation-dissipation theorem is derived connecting the memory
kernel $\tilde{K}$ and the noise $\tilde{R}$. The corresponding
modification in the Fokker-Planck equation is also discussed.

Further theoretical and experimental effort is
needed to verify the predictions of the model.
To this end, we study in detail the different physically
significant limits of the Brownian dynamics in the non
homogeneous background.
In particular it is shown that in the white noise limit
the friction coefficient in eq.(1) acquires a space-time
modulation: $\gamma_{\scriptscriptstyle 0}
\rightarrow \gamma(x,t)$, and that a
precise relation between $\gamma(x,t)$ and the stochastic potential
$\tilde{V}(x,t)$ holds.

In Sec. 3 the model is quantized via Feynman path integration.
The non homogeneous influence functional and the master equation
for the reduced density matrix of the central particle are
obtained in closed form. We discuss the semiclassical limit
and verify its consistency with the results obtained
from the classical model. The expressions obtained are compared
with those already derived in the literature in order to
analyze the consequences for the issues of decoherence and
localization.

\vspace{2cm}

{\large \bf 2. The classical model.}

\vspace{0.5cm}

{\bf 2.a Langevin equation and fluctuation-dissipation theorem.}

\vspace{0.3cm}

Our starting point is the total classical Lagrangian, eq.(4),
of a central particle $x$ in interaction with an infinite set of
harmonic oscillators $ \{ q_{n} \} $. We now consider that a non
zero external potential $V(x,t)$ acting on the central particle
is present. In this case the interpretation of $L_{osc}$ as the
Lagrangian of the whole environment does not hold and there is
no more reason to request its reflection and translation invariance.
In order to take into account the influence of the external potential
on the environment, we let the generic oscillator $q_{n}$ interact
with the external field through a coupling function $F_{n}(x(t),t)$.
The new total Lagrangian is then
\begin{equation}
L(x,\dot{x}, \{ q_{n},\dot{q}_{n} \} ,t)=
\frac{M}{2}\dot{x}^{2} - V(x,t) +
\sum_{n} \left\{ \frac{m}{2}( \dot{q}_{n}^{2} -
\omega_{n}^{2}q_{n}^{2}) + c_{n}q_{n}F_{n}(x,t) \right\} ,
\end{equation}

\noindent where the $c_{n}$'s are constants.
This Lagrangian, already considered in \cite{Moh80},
cannot lead to the correct generalized Langevin equation.
The reason is that in the homogeneous
limit, that is by choosing
the simple linear particle-oscillator
coupling $F_{n}=m\omega_{n}^{2}
x/c_{n}$, it does not reduce to Lagrangian (4), but
to the original non invariant
Feynman-Vernon Lagrangian, where the renormalization
potential $\sum_{n}m\omega_{n}^{2}x_{n}/2$ is missing.

In fact, it can be shown that
the generalized Langevin equation obtained from the Lagrangian
of eq.(12) contains a force term which is
the gradient of the nonlinear renormalization potential
$\Delta V = -\sum_{n}c_{n}F_{n}^{2}/2m\omega_{n}^{2}$.

In order to avoid the presence of this ill-defined quantity
in the equations of motion we redefine the Lagrangian by
adding the renormalization potential $-\Delta V$. It is convenient
to introduce the quantities $Q_{n}(x(t),t) = c_{n}F_{n}(x(t),t)/
m\omega_{n}^{2}$, with the physical dimensions of a
coordinate. Then Lagrangian (12) becomes
\begin{equation}
L(x,\dot{x}, \{ q_{n},\dot{q}_{n} \} ,t)=\frac{M}{2}\dot{x}^{2}
- V(x,t) + \sum_{n} \left\{ \frac{1}{2}m \dot{q}_{n}^{2}
- \frac{1}{2}m \omega_{n}^{2}\left[ q_{n} - Q_{n}(x,t)\right]^{2}
\right\} .
\end{equation}

It is to be noted that formally this Lagrangian
and the homogeneous one, eq.(4), differ only
in that the external potential $V(x,t)$ is present and
the oscillator equilibrium positions are now
represented by the functions $Q_{n}(x,t)$.

Lagrangian (13) defines the most general
oscillator model which is linear in the oscillator coordinates,
thus preserving the exact solvability of the oscillator sector,
and nonlinear in the central particle coordinates.
This Lagrangian has been considered by Zwanzig \cite{Zwa73}
and by Lindenberg and Seshadri \cite{Lind81},
in the classical regime
for the case in which the $Q_{n}$'s do not depend on time.

The equation of motion for the central
particle can be derived from Lagrangian (13)
following the same procedure applied in the homogeneous
case. Introducing the compact notations $x(t)=x$,
$x(s)=y$, and $t-s=\tau$, one
obtains the following generalized Langevin equation
\begin{equation}
M\ddot{x} + \partial_{x}V(x,t) =
- \int_{t_{0}}^{t} ds \tilde{K} (x,y,\tau)
\dot{y} - \int_{t_{0}}^{t} ds
\Phi (x,y,\tau) + \tilde{R}(x,t) \, .
\end{equation}

The first and the third terms on the right hand side of the
equation are the dissipation and fluctuation forces respectively.
The second term in the r.h.s. is an additional force, neither random
nor dissipative. Its physical origin is explained below.

The expressions for the
fluctuating force $\tilde{R}(x,t)$, the memory kernel
$\tilde{K} (x,y,\tau)$, and the deterministic
kernel $\Phi (x,y,\tau)$ are
\begin{equation}
\tilde{R}(x,t)  =  \sum_{n} m \omega_{n}^{2}
\partial_{x}Q_{n}(x,t) \Big\{ (q_{n0} - Q_{n0})\cos
[\omega_{n}(t-t_{0})] + \frac{\dot{q}_{n0}}
{\omega_{n}} \sin [\omega_{n}(t-t_{0})] \Big\} ,
\end{equation}

\begin{equation}
\tilde{K} \Big(x,y,\tau \Big) =
\sum_{n} \Big\{ \partial_{x}Q_{n}(x,t)
\partial_{y}Q_{n}(y,s) \Big\} m\omega_{n}^{2}
\cos (\omega_{n}\tau) ,~~~~~~~~~~~~~~~~~~~~~~~~~~~~~~~~
\end{equation}

\begin{equation}
\Phi \Big(x,y,\tau \Big) =
\sum_{n} \Big\{ \partial_{x}Q_{n}(x,t)
\partial_{s}Q_{n}(y,s) \Big\}
m\omega_{n}^{2} \cos (\omega_{n}\tau) ,~~~~~~~~~~~~~~~~~~~~~~~~~~~~~~~~
\end{equation}

\noindent where $Q_{n0} = Q_{n}(x(t_{0}),t_{0})$.

The oscillators $\{q_{n}\}$ of the environment are assumed to
be in thermal
equilibrium at time $t_{0}$ around their effective equilibrium
positions $Q_{n}(x,t)$ according to a canonical distribution,
in complete analogy with the homogeneous case:
\begin{equation}
\begin{array}{lll}
\langle q_{i0} - Q_{i0} \rangle & = & 0 \, , \\
&& \\
\langle \dot{q}_{i0} \rangle & = & 0 \, , \\
&& \\
m\omega_{i}\omega_{j} \langle (q_{i0}-Q_{i0})
(q_{j0}-Q_{j0}) \rangle & = & k_{B}T \delta_{ij} \, , \\
&& \\
m\langle \dot{q}_{i0}\dot{q}_{j0} \rangle & = & k_{B}T
\delta_{ij} \, , \\
&& \\
\langle \dot{q}_{i0}(q_{j0} - Q_{j0}) \rangle & = & 0 \, .
\end{array}
\end{equation}

Comparing the definition (15) of $\tilde{R}$
with the above statistical conditions on the oscillators, one
can prove that the fluctuating force is a Gaussian
stochastic process with zero mean
\begin{equation}
\left\langle \tilde{R}(x,t) \right\rangle = 0,
\end{equation}

\noindent and covariance
\begin{equation}
\left\langle \tilde{R}(x,t) \cdot \tilde{R}(y,s) \right\rangle
= k_{B}T\tilde{K}(x,y,\tau) \, .
\end{equation}

It is remarkable that a fluctuation-dissipation theorem
holds even in this non homogeneous
generalization of the oscillator model.
Furthermore, no restriction has been assumed on the different
time scales of the noise and of the external potential.
{}From the definition (15) and the statistical properties (19)-(20),
it follows that the fluctuating force $\tilde{R}$
is a colored multiplicative process, in general not
factorizable in the product of
a time-dependent and of a space-dependent part.

If the functions $Q_{n}$ have no
explicit time dependence, then from eq.(17) it follows that
the last force term in the Langevin equation is absent.
On the other hand, if the functions $Q_{n}$ do explicitly
depend on time but not on $x(t)$, then from eq.(16) it follows
that the dissipative force is absent,
and we are left with a purely time-dependent force,
a case already considered by Feynman and Vernon \cite{Fey63}.
Finally, the case of homogeneous noise is regained as $V=0$ and
$Q_{n}(x,t) \equiv x$, $\forall n$, and eqs.(13)-(16) and
(18)-(20) go over into the corresponding eqs.(4),(10),(7),(8)
and (5),(2),(9).

The Langevin equation obtained above is very general
because, as we earlier remarked,
the noise $\tilde{R}$ and the correlation function
$\tilde{K}$ do not factorize. However, when
all the functions $Q_{n}$ are equal to
the same function $Q(x,t)$, the equation
simplifies into:
\begin{equation}
M\ddot{x} + \partial_{x}V(x,t) =
\partial_{x}Q(x,t) \Bigg\{ R(t) -
\int_{t_{0}}^{t} K(\tau)\Big( \partial_{s}Q(y,s)
+ \dot{y}\partial_{y}Q(y,s) \Big) ds \Bigg\} \; ,
\end{equation}

\noindent where $R(t)$ and $K(\tau)$ are those of the homogeneous
case, given by eqs.(7)-(8), and, as before, $y$ stands for $x(s)$.
As a consequence, the generalized fluctuation-dissipation theorem
(20) reduces to the homogeneous form (9).

Inspection of eq.(21)
shows that the effect of the new oscillator equilibrium position
amounts to a space-time modulation of the fluctuating force, which
can be rewritten as
\begin{equation}
\tilde{R}(x,t) = \partial_{x}\Big( R(t)Q(x,t) \Big)
= -\partial_{x}\tilde{V}(x,t) \; .
\end{equation}

\noindent We are thus naturally led to the interpretation of
$-Q(x,t)$ as a deterministic modulation of the fluctuating
potential $\tilde{V}$.
It is now clear that the last force term on the right hand side
of eq.(21) derives from the time modulation alone of the stochastic
force. Namely, the model predicts that if there is a time modulation
of the noise, then, besides the usual dissipative and random terms,
a new (non dissipative) force appears.

If the functions $Q_{n}(x,t)$ are different from each other, then the
random force $\tilde{R}(x,t)$ in eq.(14) can still be written as the
gradient of a stochastic potential $\tilde{V}(x,t)$,
given by the expression
\begin{equation}
\tilde{V}(x,t) = - \sum_{n}m\omega_{n}^{2}Q_{n}(x,t)
\left\{ (q_{n0} - Q_{n0})\cos [\omega_{n}(t-t_{0})] +
\frac{\dot{q}_{n0}}{\omega_{n}}\sin [\omega_{n}(t-t_{0})] \right\} .
\end{equation}

In this case we preserve the same interpretations of the functions
$Q_{n}$. The above relation can be seen as a particular
Fourier expansion
of the random potential, in which every Fourier component
is modulated in space and time by the
functions $Q_{n}(x,t)$.

In the white noise limit
$K(\tau) \rightarrow 2M\gamma_{\scriptscriptstyle 0}
\delta (\tau)$ and eq.(21) reduces to
\begin{equation}
M\ddot{x} + \partial_{x}V(x,t) = - M \gamma(x,t)\dot{x} +
R(t)\partial_{x}Q(x,t)
- M\gamma_{\scriptscriptstyle 0} \partial_{x}Q(x,t)
\partial_{t}Q(x,t) \; ,
\end{equation}

\noindent where we have defined the effective friction
coefficient as
\begin{equation}
\gamma(x,t) = \gamma_{\scriptscriptstyle 0} \Big(
\partial_{x}Q(x,t) \Big)^{2} \; .
\end{equation}

We see that the nonlinear coupling to the linear bath
modifies the white-noise Langevin equation by introducing
a nonlinear dissipative force, a multiplicative
noise term, and a deterministic force due to the
time-variations of the coupling.
The friction coefficient $\gamma_{\scriptscriptstyle 0}$ is
modulated by the square of the function that modulates the
white noise $R(t)$, as expected from the
fluctuation-dissipation theorem (20). For
$Q_{n}(x,t)=Q(x,t)$, the following relation holds:
\begin{equation}
\tilde{R}(x,t) = -\partial_{x}\tilde{V}(x,t)
= R(t)\partial_{x}Q(x,t) =
\sqrt{\frac{\gamma(x,t)}{\gamma_{\scriptscriptstyle 0}}} R(t).
\end{equation}

It is instructive to analyze the overdamped regime as well. In this
case however, one must be aware of the fact that
the very existence of the overdamped limit
rests on the hypothesis that the scale
of the time-variations of $Q(x,t)$ is
slower than the relaxation time $1/\gamma_{\scriptscriptstyle 0}$.

Assuming that this constraint is satisfied, in the overdamped
regime one has that the
function modulating the fluctuation goes into its inverse:
\begin{equation}
\dot{x} =
-\frac{1}{\partial_{x} Q} \left( \frac{\partial_{x} V}{M
\gamma_{\scriptscriptstyle 0} \partial_{x} Q} -
\frac{R}{M\gamma_{\scriptscriptstyle 0}}
+ \partial_{t} Q \right) \, .
\end{equation}

When $\partial_{t} Q=0$, according to the above
expression the particle moves under the influence of
an effective force $-\partial_{x}V/{\partial_{x}Q}^{2}$.
Compared to the homogeneous case, the modulation of the noise can
strongly modify the intensity of the external force in the overdamped
limit but cannot change its direction. However, when $\partial_{t} Q
\neq 0$, also the sign of the effective force can be reversed if
the scale of time-variations of $Q(x,t)$ is small enough.

Finally, if $Q$ does not explicitly depend on time, eq.(27) suggests
a simple method to determine $\partial_{x} Q$ and so the nonlinear
coupling Q as well. From a measurement of the
velocity $\dot{x}$ and the knowledge of the external potential
$V(x,t)$, it is possible to infer the non homogeneous nature of
the environment and to determine $\partial_{x} Q$
given that, on average, ${\partial_{x} Q}^{2} = -
\partial_{x} V/M\gamma_{\scriptscriptstyle 0}\dot{x}$.

\vspace{0.5cm}

{\bf 2b. Fokker-Planck equation.}

\vspace{0.3cm}

In the case of white noise the Langevin equation
\begin{equation}
M\ddot{x}(t) + \partial_{x}V(x,t)
= - M\gamma_{\scriptscriptstyle 0}\dot{x}(t) + R(t) \, ,
\end{equation}

\noindent is equivalent to the Fokker-Planck equation
\begin{equation}
\partial_{t}W = -pM^{-1}\partial_{x}W
+ \partial_{p}\Big( W\partial_{x}V +
\gamma_{\scriptscriptstyle 0}pW \Big)
+ M\gamma_{\scriptscriptstyle 0}\beta^{-1}
\partial_{p}^{2}W
\end{equation}

\noindent for the phase-space probability
density $W(x,p,t)$, where $p$ is the momentum of the particle
and $\beta^{-1} = k_{B}T$.

In the overdamped regime the above equation reduces to the
diffusion equation for the configurational probability density
$P(x,t) = \int dp W(x,p,t)$:
\begin{equation}
\partial_{t}P =
\Big( M\gamma_{\scriptscriptstyle 0} \Big)^{-1} \Big[
\partial_{x} \Big( P\partial_{x}V \Big)
+ \beta^{-1}\partial_{x}^{2}P \Big] \, .
\end{equation}

Considering now the case of non
homogeneous noise, the Fokker-Planck equation equivalent to
the white noise Langevin equation (24) is
\begin{equation}
\partial_{t}W = -pM^{-1}\partial_{x}W
+ \partial_{p}\Big[ \Big(
\gamma p + M\gamma_{\scriptscriptstyle 0}\partial_{x}Q
\partial_{t}Q + \partial_{x}V \Big) W \Big]
+ M\gamma\beta^{-1}
\partial_{p}^{2}W \, .
\end{equation}

It is to be noted that since the same function $\gamma(x,t)$
appears both in the drift and in the diffusion term, the
equilibrium (stationary) solution of eq.(31), if $Q$ and $V$
do not depend on time and if such a solution exists, is just
the Maxwell-Boltzmann distribution $W_{eq.}(x,p) \propto
\exp \left[ -\beta \left( p^{2}/2 + V(x) \right)
\right] $.

In the overdamped regime, one obtains the diffusion equation
for the configurational probability density $P(x,t)$:
\begin{equation}
\partial_{t}P = \partial_{x} \left[ \left(
\frac{\partial_{x}V}{M\gamma} + \frac{\beta^{-1}}{M
\gamma_{\scriptscriptstyle 0}}
\frac{\partial_{t}Q}{\partial_{x}Q} -
\frac{\partial_{x}^{2}Q}{(\partial_{x}Q)^{3}} \right) P \right] +
\frac{\beta^{-1}}{M\gamma_{\scriptscriptstyle 0}} \Big(
\partial_{x}\Big[ (\partial_{x}Q)^{-2} \Big] \Big)
\partial_{x}P \, .
\end{equation}

This equation can be used to study specific examples of nonlinear
couplings with given external potentials.
It is easy to see that the modifications induced by the
non homogeneous terms may lead to interesting non perturbative
effects.
We shall present in a
forthcoming paper a detailed treatment of the non homogeneous
Brownian dynamics in bistable and in periodic potentials.

\vspace{0.6cm}

{\large \bf 3. The quantum model}

\vspace{0.4cm}

{\bf 3.a Influence functional.}

\vspace{0.2cm}

We shall now discuss the quantum version of the generalized
oscillator model, by quantizing the Lagrangian (13) via Feynman
path integration, and derive (see Sec.3.b) the relevant evolution
equations for the reduced density matrix $\rho$ of the central
Brownian particle.

We begin by considering the action $S[x,q]$ of
a system formed by a central particle $x$ and
just one oscillator $q$ of natural frequency $\omega / 2\pi$:
\begin{eqnarray}
& & S[x,q] = S_{0}[x] + S_{osc}[q,x] = \nonumber \\
& & \\
& & = \int_{t_{a}}^{t_{b}} dt \left\{ \frac{M}{2} \dot{x}^{2} -
V(x,t) \right\} + \int_{t_{a}}^{t_{b}} dt \frac{m}{2} \left\{
\dot{q}^{2} - \omega^{2}(q-Q(x,t))^{2} \right\} \, , \nonumber
\end{eqnarray}

\noindent where $t_{a}$ and $t_{b}$ are two arbitrary initial
and final times, respectively. The functions $S_{0}$ and $S_{osc}$
are respectively the actions of the isolated particle and of the
oscillator, while $Q(x,t)$ is an arbitrary space-time
function.

For the sake of clarity,
we shall assume here and in the following
that the initial density matrix of the total system factorizes
in the product of the density matrices of the two subsystems.
In the coordinate representation, denoting by
$f_{i}$ and $f'_{i}$ two different values of an observable
quantity $f(x,t)$ at time $t_{i}$, the former statement reads
\begin{equation}
\rho (x_{a},x'_{a},q_{a},q'_{a},t_{a}) = \rho (x_{a},x'_{a},t_{a})
\cdot \rho_{\ast}(q_{a} - \bar{Q}_{a} , q'_{a} - \bar{Q}_{a}) .
\end{equation}

\noindent Some remarks should be made about the form of this
density matrix.
Equation (34) describes the result of a measurement in which
the state of the central particle has not been affected by the
environment (\cite{Gra88}, \cite{Sch87}, \cite{Pat94}).
The density
matrix $\rho (x_{a},x'_{a},t_{a})$ represents the
initial state of the central particle, and $\rho_{\ast}(q_{a}
- \bar{Q}_{a} , q'_{a} - \bar{Q}_{a})$ is the equilibrium
density matrix of a harmonic oscillator at an inverse
temperature $\beta = 1/k_{B}T$ with its equilibrium
position at $q_{a} = \bar{Q}_{a}$.
It has the expression
\begin{eqnarray}
\rho_{\ast}(q_{a}-\bar{Q}_{a},q'_{a}-\bar{Q}_{a}) & = &
F(\beta)\exp
\Bigg\{ - \frac{m\omega}{2\hbar}
\Bigg[ ((q_{a} - \bar{Q}_{a})^{2} +
(q'_{a} - \bar{Q}_{a})^{2})\coth
(\beta \hbar \omega) \nonumber \\
&& \nonumber \\
& - & \frac{2(q_{a} - \bar{Q}_{a})
(q'_{a} - \bar{Q}_{a})}
{\sinh (\beta \hbar \omega)} \Bigg] \Bigg\} \; ,
\end{eqnarray}

\noindent where $F(\beta) = \sqrt{m\omega/ \pi \hbar \coth
(\beta \hbar \omega/2)}$ is a normalization factor such that
$\int dq \rho_{\ast}(q,q) = 1$.

The oscillator density matrix only depends
on $q_{a}$ and $q'_{a}$ through the differences $q_{a} -
\bar{Q}_{a}$ and $q'_{a} - \bar{Q}_{a}$, where $\bar{Q}_{a}$,
in analogy with the case of homogeneous dissipation \cite{Pat94},
can be shown to be given by
\begin{equation}
\bar{Q}_{a} = \frac{Q(x_{a},t_{a}) + Q(x'_{a},t_{a})}{2} .
\end{equation}

This expression accounts for the fact that, compared to the
homogeneous case, in which
$\bar{Q}_{a} = (x_{a} + x'_{a})/2$,
the equilibrium position of the oscillator
at time $t$ has been shifted from $x(t)$ to $Q(x(t),t)$.

The density matrix evolves in time from $t=t_{a}$ to
$t=t_{b}$ through the action of the total density matrix
propagator $J(b|a) \equiv J(x_{b},x'_{b},q_{b},q'_{b},t_{b}|x_{a},
x'_{a},q_{a},q'_{a},t_{a})$:
\begin{equation}
\rho(x_{b},x'_{b},q_{b},q'_{b},t_{b}) = \int dx_{a} dx'_{a}
dq_{a} dq'_{a} J(b|a)
\rho (x_{a},x'_{a},q_{a},q'_{a},t_{a}) .
\end{equation}

The total propagator $J(b|a)$ is simply given by the
product of the two propagators of the total wave function:
\begin{equation}
J(b|a) = \int_{x_{a}}^{x_{b}} {\cal{D}} x
\int_{x'_{a}}^{x'_{b}} {\cal{D}} x' \int_{q_{a}}^{q_{b}}
{\cal{D}} q \int_{q'_{a}}^{q'_{b}} {\cal{D}} q' \exp
\left\{ \frac{i}{\hbar}(S[x,q] - S[x',q']) \right\} \, ,
\end{equation}

\noindent where $S[x,q]$ is given by eq.(33). The reduced
density matrix of the central particle at $t=t_{b}$ is
obtained by integrating the total density matrix in the
oscillator coordinate,
\begin{equation}
\rho(x_{b},x'_{b},t_{b}) = \int dq_{b} \rho (x_{b},x'_{b},
q_{b},q'_{b},t_{b}) .
\end{equation}

Following the same procedure as in the homogeneous case
(\cite{Fey63}, \cite{Cal83}), the time evolution
law of the reduced density matrix is
\begin{equation}
\rho(x_{b},x'_{b},t_{b}) = \int dx_{a} dx'_{a}
J_{eff}(x_{b},x'_{b},t_{b}|x_{a},x'_{a},t_{a})
\rho(x_{a},x'_{a},t_{a}) ,
\end{equation}

\noindent where the effective propagator $J_{eff}$ can be written as
\begin{equation}
J_{eff}(x_{b},x'_{b},t_{b}|x_{a},x'_{a},t_{a}) =
\int_{x_{a}}^{x_{b}} {\cal{D}} x \int_{x'_{a}}^{x'_{b}}
{\cal{D}} x' \exp \left( \frac{i}{\hbar} {\cal{S}} [x,x']
\right) .
\end{equation}

Here we have introduced the effective action
\begin{equation}
{\cal{S}} [x,x'] = S_{0}[x] - S_{0}[x'] + \Phi [x,x'] ,
\end{equation}

\noindent where $\Phi[x,x']$ denotes the influence phase \cite{Fey63},
and must not be confused with the deterministic kernel $\Phi (
x,y,\tau )$ defined by eq.(17).
Introducing, as before, the compact notation
$x'=x'(t)$, $x=x(t)$, $y'=x'(s)$, $y=x(s)$, and
$t-s=\tau$, the expression for the influence phase reads
\begin{eqnarray}
\Phi[x,x'] & = &
\frac{m\omega^{2}}{2}\int_{t_{a}}^{t_{b}} dt
\int_{t_{a}}^{t} ds \Big[ Q(x',t) - Q(x,t) \Big]
\Big\{ \partial_{s}Q(y,s) +
\partial_{s}Q(y',s) \Big\} \cos ( \omega\tau ) \nonumber \\
&& \nonumber \\
& + & \frac{m\omega^{2}}{2}\int_{t_{a}}^{t_{b}} dt
\int_{t_{a}}^{t} ds \Big[ Q(x',t) - Q(x,t) \Big]
\Big\{ \dot{y}\partial_{y}Q(y,s) +
\dot{y}'\partial_{y'}Q(y',s) \Big\}
\cos ( \omega\tau ) \\
&& \nonumber \\
& + & i\frac{m\omega^{3}}{2}\coth (\beta \hbar \omega /2)
\int_{t_{a}}^{t_{b}} dt
\int_{t_{a}}^{t} ds \Big[ Q(x',t) - Q(x,t) \Big]
\Big\{ Q(y',s) - Q(y, s) \Big\}
\cos ( \omega\tau ) \, . \nonumber
\end{eqnarray}

Because of the dynamical
and statistical independence of the oscillators, in the
case of an environment with infinite degrees of freedom the influence
phase can be calculated by simply summing up the influence phases of
all the oscillators.

If all the $Q_{n}$'s are
taken to be equal to a certain function $Q$,
the influence phase assumes the form
\begin{eqnarray}
\Phi[x,x'] & = & \frac{1}{2}\int_{t_{a}}^{t_{b}} dt
\int_{t_{a}}^{t} ds \Big[ Q(x',t) - Q(x,t) \Big]
\Big\{ \partial_{s}Q(y,s)
+ \partial_{s}Q(y',s) \Big\} K(\tau) \nonumber \\
&& \nonumber \\
& + & \frac{1}{2}\int_{t_{a}}^{t_{b}} dt
\int_{t_{a}}^{t} ds \Big[ Q(x',t) - Q(x,t) \Big]
\Big\{ \dot{y}\partial_{y}Q(y,s)
+ \dot{y}'\partial_{y'}Q(y',s) \Big\} K(\tau) \nonumber \\
&& \nonumber \\
& + & i\int_{t_{a}}^{t_{b}} dt \int_{t_{a}}^{t} ds
\Big[ Q(x',t) - Q(x,t) \Big]
\Big\{ Q(y',s) - Q(y,s) \Big\} \alpha(\tau) \, ,
\end{eqnarray}

\noindent where
\begin{equation}
\alpha(\tau) = \frac{1}{2}\sum_{n}m\omega_{n}^{3}\coth (\beta
\hbar\omega_{n}/2)\cos(\omega_{n}\tau) \, ,
\end{equation}

\noindent and the function $K(\tau)$ is the same memory kernel
defined in the classical homogeneous problem, eq.(8).

Given the expressions (41) and (44), the corresponding
evolution law (37) for the reduced density matrix is
completely determined and provides the general description of
the dynamics of a quantum Brownian particle acted upon by a
multiplicative stochastic non homogeneous potential
$\tilde{V}(x,t)= -Q(x,t)R(t)$.

\vspace{0.8cm}

{\bf 3.b Master equation and decoherence.}

\vspace{0.3cm}

Starting from the functional integral formulation, we can
also write the master equation for the reduced density matrix,
which provides an alternative description of quantum Brownian
motion.

A differential evolution equation for the density matrix $\rho(x,x',t)$
exists if the effective Lagrangian corresponding to
the effective action (42) is local in time. This happens
in the white noise limit, in which the influence phase reads
\begin{eqnarray}
\Phi[x,x'] & = & M\gamma_{\scriptscriptstyle 0}\int_{t_{a}}^{t_{b}} dt
\Big( Q(x',t) - Q(x,t) \Big) \Bigg( \dot{Q}(x,t) +
\dot{Q}(x',t) \Bigg) \nonumber \\
& & \nonumber \\
& + & M\gamma_{\scriptscriptstyle 0} \int_{t_{a}}^{t_{b}} dt
\Big( Q(x',t) - Q(x,t) \Big) \Bigg(
\dot{x}\partial_{x}Q(x,t)
+ \dot{x}'\partial_{x'}Q(x',t) \Bigg) \nonumber \\
& & \nonumber \\
& + & i \frac{M\gamma_{\scriptscriptstyle 0}}{\beta\hbar}
\int_{t_{a}}^{t_{b}} dt
\Big( Q(x',t) - Q(x,t) \Big)^{2} \, .
\end{eqnarray}

The white noise master equation can now be derived
following the same procedure applied
by Caldeira and Leggett in the linear
homogeneous case \cite{Cal83}.
Starting from the effective action (42) with the nonlinear
influence phase (46) we obtain
\begin{eqnarray}
\frac{\partial \rho }{\partial t} & = &
\frac{i\hbar}{2M} \left( \frac{\partial^{2}}{\partial x^{2}}
- \frac{\partial^{2}}{\partial x'^{2}} \right)\rho -
\frac{i}{\hbar} \Bigg( V(x',t) - V(x,t) \Bigg)\rho \nonumber \\
&& \nonumber \\
& - & \frac{iM}{\gamma_{\scriptscriptstyle 0}\hbar} \Bigg(
Q(x',t) - Q(x,t) \Bigg) \Bigg( \frac{\partial Q(x,t)}{\partial t}
+ \frac{\partial Q(x',t)}{\partial t} \Bigg) \rho \nonumber \\
&& \nonumber \\
& - & \frac{\gamma_{\scriptscriptstyle 0}}{2} \Bigg( Q(x,t)
- Q(x',t) \Bigg)
\left( \frac{\partial Q(x,t)}{\partial x}
\frac{\partial}{\partial x} - \frac{\partial Q(x',t)}{\partial x'}
\frac{\partial}{\partial x'} \right) \rho \nonumber \\
&& \nonumber \\
& - & \frac{2M\gamma_{\scriptscriptstyle 0}}{\beta \hbar^{2}}
\Bigg( Q(x,t) - Q(x',t) \Bigg)^{2} \rho \; .
\end{eqnarray}

The first two terms on the right hand side of eq.(47) represent
the mechanical evolution without bath-particle interaction.
The fourth one describes dissipation and the last one the random
fluctuations which destroy the quantum coherence. The third term
is present only when the bath-particle coupling is time-dependent,
and it can completely modify the usual decoherence patterns as
we will clarify below.

The Caldeira-Leggett master equation in a homogeneous environment
is recovered for $Q \equiv x$, that is for $\partial_{x}Q \equiv 1$.

In the classical limit eq.(47) reduces to
the Fokker-Planck dynamics discussed in Sec.2b.
This can be best understood by transforming
the master equation (47) for the density matrix
in the evolution equation for the associated Wigner function.
The task is achieved by introducing the ``relative
coordinate" $Y=x'-x$ and the ``center of mass coordinate"
$X=(x+x')/2$ and by Fourier transforming eq.(47)
respect to the $Y$ variable.
In this way, one obtains the evolution equation in phase space
for the Wigner function $\hat{W}(x,p,t)$ defined as
\begin{equation}
\hat{W}(x,p,t) = \int dY \rho \Bigg( X - \frac{Y}{2}, X +
\frac{Y}{2} \Bigg)
e^{ipY/\hbar} \; .
\end{equation}

It is then easy to prove that, in the limit of small $\hbar$,
the evolution
equation for the Wigner function $\hat{W}(x,p,t)$ reduces to the
Fokker-Planck equation (31) for the classical probability density
$W(x,p,t)$, where $p$ is the momentum of the particle.

The quantum model provides very interesting information
concerning the dynamics of decoherence and the transition from
quantum to classical.
We consider in particular the coefficient $g(x,x',t)$ of
$\rho(x,x',t)$ in
the last term of the master equation (47),
\begin{equation}
g(x,x',t) = \frac{2M\gamma_{\scriptscriptstyle 0}}{\beta \hbar^{2}}
\Bigg( Q(x,t) - Q(x',t) \Bigg)^{2} ,
\end{equation}

\noindent which is always positive and
represents the space-time dependent decay rate
of the off-diagonal elements in the density matrix;
it becomes very large
in the semiclassical limit, since it is proportional
to $\hbar^{-2}$ \cite{Zur93}, and
it reduces to a quadratic function
of the difference $|x-x'|$ in the homogeneous case. In the
non homogeneous case this term, as well as the dissipative one,
loses its invariance properties under reflection and translation,
because of its dependence on the noise modulating function $Q(x,t)$.

When the function $Q$ depends explicitly on time, eq.(47) describes
a decoherence process with multiple nonequilibrium scales
due to time-modulation of the noise. Explicitely, the different
scales are represented by the decoherence rate $g$, the
damping rate $\gamma$, the external potential $V(x,t)$,
and the modulating function $Q(x,t)$.

Overall, these nonequilibrium effects are expressed by
the third term on the r.h.s. of the master equation; formally, it
is similar to the terms derived from a potential, that do not
involve space-derivatives, with the crucial difference that it
cannot be written as a difference of the type $V(x',t)-
V(x,t)$, and so its effect is not equivalent, as in the
classical case, to adding a suitable external potential.
When the physical situation is such that the bath-particle coupling
is strongly time-dependent, possibly with very fast variations, this
term yields a fastly-oscillating contribution in the solution and
so can reduce or even inhibit the decoherence process.

Generalizations of eq.(47) may be considered in order
to take into account other effects such as
bath-particle interactions nonlinear in the oscillator coordinates
or the finite size of the bath.

In the first case the decoherence rate $g(x,x',t)$ does not
depend quadratically on the difference $Q(x,t)-Q(x',t)$
(or on the difference $x-x'$ in the homogeneous case),
because such a dependence is a direct consequence of
the oscillator Lagrangian being quadratic.
In the second case quantum interference is suppressed only as long
as the system is confined inside the thermal bath,
which is assumed to possess a finite scale
$\xi$, of the order of its linear dimensions.

The variable $Y = x - x'$ is conjugate to the variable $p$
(see the definition of the Wigner function)
which in the semiclassical limit reduces to the momentum variable,
and actually represents the quantum uncertainty on the position of the
Brownian particle \cite{Schmid82}. We then expect expression (48)
to be replaced by a function
$g(x,x',t)$ which becomes negligible for $|x - x'| \gg \xi $.

A finite value of (48)
for $|x - x'| \gg \xi $, both in the
homogeneous and in the non homogeneous models,
is possible only if the bath is
assumed to be infinitely extended. This
point clarifies some questions recently raised
\cite{Gallis93} concerning the decoherence process.

In this paper we generalized the homogeneous model of
classical and quantum dissipation
in order to include general bath-particle interactions
nonlinear in the particle coordinates.
The linear-dissipation theory emerges as a particular case
of the model.
Moreover, we have shown that the nonlinear bath-particle coupling
leads to the appearance of non homogeneous random
and dissipative forces and of a further deterministic
time-dependent force, in the Langevin equation at the classical
level, and in the master equation for the quantum case.
We illustrated how this yields new interesting consequences
both in the classical and in the quantum regime.

\end{document}